\documentclass[conference]{IEEEtran}

\usepackage{graphicx}
\newtheorem{theorem}{Theorem}
\newtheorem{lemma}{Lemma}
\newtheorem{definition}{Definition}
\newtheorem{example}{Example}
\usepackage[caption=false,font=footnotesize]{subfig}
\usepackage{amssymb}

\begin{document}

\title{Secure rate-adaptive reconciliation}

\author{
  \IEEEauthorblockN{
    David~Elkouss,
    Jes\'us~Mart\'inez-Mateo
    and~Vicente~Mart\'in
  }\\
  \IEEEauthorblockA{
    Research group on Quantum Information and Computation\\
    Universidad Polit\'ecnica de Madrid (UPM)\\
    Campus de Montegancedo, 28660 Boadilla del Monte (Madrid), Spain\\
    e-mail: \{delkouss, jmartinez, vicente\}@fi.upm.es
  }
}

\maketitle

\begin{abstract}

We consider in this paper the problem of information reconciliation in the context of secret key agreement between two legitimate parties, Alice and Bob. Beginning the discussion with the secret key agreement model introduced by Ahlswede and Csisz\'ar, the channel-type model with wiretapper, we study a protocol based on error correcting codes. The protocol can be adapted to changes in the communication channel extending the original source. The efficiency of the reconciliation is only limited by the quality of the code and, while transmitting more information than needed to reconcile Alice's and Bob's sequences, it does not reveal any more information on the original source than an ad-hoc code would have revealed.

\end{abstract}

\section{Introduction}

\begin{figure}[h!]
\includegraphics[width=1.0\linewidth]{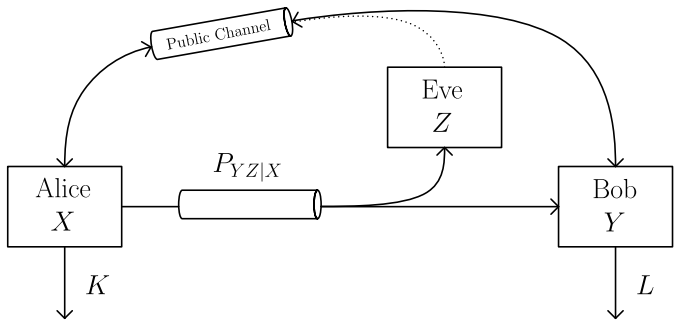}
\caption{Ahlswede and Csisz\'ar's model CW.}
\label{fig:model-cw}
\end{figure}

Lets start by considering the channel-type model with wiretapper (CW) for secret key agreement introduced by Ahlswede and Csisz\'ar~\cite{Ahlswede_93} as shown in Fig.~\ref{fig:model-cw}. In this model a legitimate party, Bob, and an eavesdropper, Eve, are both connected to another legitimate party, Alice, through a discrete memoryless channel (DMC). Alice generates a discrete sequence of $n$ values, $X^n$, while Bob and Eve observe the correlated outputs, $Y^n$ and $Z^n$ respectively, obtained after the transmission of $X^n$ over the DMC. Both outputs are characterised by transition probability $P_{Y,Z|X}$, with each component of the sequences being the outcome of an independent use of the channel. Alice and Bob have also access to a public but authenticated channel used to distill a shared secret key from their correlated sequences. Public and authenticated means in this context that Eve has noiseless access to the information exchanged through the channel, but she is not able to tap the channel without being noticed. Therefore the integrity of the messages on the public channel is guaranteed.

Protocols that distill a secret key usually divide the distillation process in two different phases. In the first one, known as information reconciliation or simply reconciliation, Alice and Bob exchange redundant information over the public channel in order to eliminate any discrepancy in their correlated sequences, $X^n$ and $Y^n$ respectively. At the end of the reconciliation phase both parties have agreed on a shared secret string $\chi$, though in many cases $\chi=X^n$. On the second phase, known as privacy amplification, Alice and Bob shrink their strings in order to wipe any information of the previously shared key that the eavesdropper could have on $\chi$ through $Z^n$ or through any communication over the public channel with information about the strings. This construction allows to split the secret key distillation process into two easier problems. This division is not necessarily suboptimal and, as it is shown in section~\ref{sec:security}, under certain conditions Alice and Bob can achieve the maximal secret key rate.

The paper is organised as follows: Section~\ref{sec:problem} includes a review of the information reconciliation problem linking it with secret key agreement. Section~\ref{sec:protocol} describes an information reconciliation protocol over an extended string; this protocol uses Wyner's coset scheme with Low-Density Parity-Check (LDPC) codes~\cite{Richardson_01} and can achieve an efficiency as close to its optimum value as allowed by the quality of the code. In section~\ref{sec:security} it is proved that the proposed protocol does not reveal any more information on $X$ than an adapted solution for string $X$ would reveal. And finally, section~\ref{sec:num} analyses the performance of this protocol in a practical scenario.

\section{Problem Statement}
\label{sec:problem}
Secret key distillation process is usually divided into privacy amplification and information reconciliation. This section defines the meaning of secret key in the context of this paper. Then privacy amplification and information reconciliation are introduced and linked. The objective is to highlight the influence of efficient reconciliation in the achievable secret key rate.

\subsection{Secret Key Agreement}

Alice, Bob and Eve hold $n$-length sequences, $X^n$, $Y^n$ and $Z^n$ respectively, with each component of the sequences being characterised by $P_{Y,Z|X}$.

Let $\phi_i$ denote the message that Alice sends over the public channel in its $i$-th use, and $\psi_i$ denote the message that Bob sends in his $i$-th use of the channel. Both sets of messages or communications, $\phi$ and $\psi$, are respectively known as forward and backward transmissions. Depending on the protocol any individual message or even $\phi$ or $\psi$ might be null. The former case, $\phi \in \emptyset$, is known as direct reconciliation while the latter, $\psi \in \emptyset$, is known as reverse reconciliation. After $k$ uses of the public channel, i.e. after the exchange of the set of Alice's first $k$ messages, $\phi^k$, and Bob's, $\psi^k$ messages, Alice and Bob estimate their shared keys to be $K$ and $L$ respectively by using an agreed protocol.

\begin{definition}

A \textit{strong} secret key rate $S$ is achievable if there exist $(\phi^k, \psi^k)$ that for large enough $n$ and for every $\epsilon > 0$ that meets simultaneously the following restrictions \cite{Liang_09}:

\begin{equation}
\Pr [K \neq L] < \epsilon
\end{equation}

\begin{equation}
I(\phi^k, \psi^k, Z^n; K) < \epsilon 
\end{equation}

\begin{equation}
H(K) > n\cdot S - \epsilon 
\end{equation}

\begin{equation}
\log |K| < H(K) + \epsilon 
\end{equation}

\end{definition}

\noindent where $H(\cdot )$ stands for Shannon's entropy, while $I(\cdot ;\cdot )$ stands for Shannon's mutual information. This definition of secret key rate is strong compared to previous definitions in which the convergence of the conditions was asymptotic and not absolute. In \cite{Maurer_00} it is shown that both sets of conditions share the same bounds for secret key generation.

Henceforth the superindex indicating length is dropped to reduce the notation, the length of the variable or string should be clear from the context, whenever in doubt we clarify the value that the superindex is taking. 

The largest achievable secret rate $S$ is upper bounded by the secret key capacity, $C_S$, which if only forward communications are allowed is defined by~\cite{Ahlswede_93}:

\begin{equation}
C_{S_f}=\max_{P_{U,X}}[I(U;Y)-I(U;Z)]
\end{equation}

\noindent where $U$ is an auxiliary random variable that forms the Markov chain $U\rightarrow X\rightarrow YZ$. 

It should be noticed that $C_{S_{f}}$ is a lower bound of $C_S$ if two way communications are allowed~\cite{Vanassche_05}. A case of special interest arises when $U$ cannot be maximised or $X$ cannot be manipulated by Alice, an example of this situation is a Quantum Key Distribution (QKD) protocol fixing $X$ \cite{Bennett_84}. In this case, taking into account the restrictions, the previous result allows Alice and Bob to achieve at least a secret rate of

\begin{equation}
I(X;Y) - I(X;Z) = H(X|Z) - H(X|Y)
\label{eq:srate}
\end{equation}

\noindent where $H(X|Y)$ and $H(X|Z)$ are the Shannon conditional entropy. 

\subsection{Privacy Amplification and Information Reconciliation}
The problem of privacy amplification ---how to reduce $I(X;Z)$, the knowledge that Eve might have gathered during the process--- has been widely studied. Some of the results on privacy amplification are based on the use of universal families of hash functions~\cite{Bennett_95}, however in this work we use extractors~\cite{Nisan_93}, proposed by Maurer and Wolf for privacy amplification~\cite{Maurer_00}, as they allow to prove the strong secret key rate bounds. An extractor is a function that, with a small amount of random bits acting as catalyst, obtains a number of almost uniformly distributed random bits from a source. The main result, that we develop in section~\ref{sec:security}, states that given an upper bound on the information the eavesdropper has, Alice and Bob can extract a smaller and highly secret key. The length of the new key is a function of a security parameter and of the upper bound on Eve's information, which in turn depends on: the information that Eve gathers on the private channel and the information that Eve gathers in the information reconciliation phase, directly linking privacy amplification with information reconciliation. 

\begin{figure}
\centering
\includegraphics[width=\linewidth]{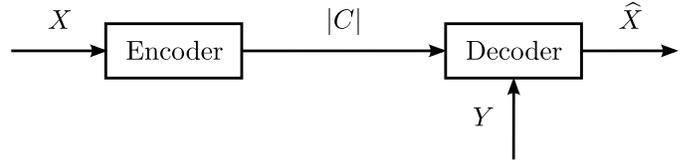}
\caption{Source coding with side information.}
\label{fig:side-information}
\end{figure}

Information reconciliation in the context of secret key agreement is also a well known problem. Once it has been separated from privacy amplification, the problem is reduced to one of Slepian-Wolf coding~\cite{Slepian_73} (see Fig.~\ref{fig:side-information}). Given a source $X$, it is sufficient a rate $R \geq H(X)$ to losslessly encode $X$, and given two sources $X$ and $Y$ to an individual encoding terminal it is sufficient with $R \geq H(X,Y)$. The surprising result by Slepian and Wolf states that even for separate encoding $R \geq H(X,Y)$ is enough~\cite{Slepian_73} and, of particular interest in information reconciliation, that it is also enough for Alice to encode her source $X$ with $R \geq H(X|Y)$ in order to allow Bob infer $X$.

Wyner's coset scheme is a good solution for the compression of binary sources with side information~\cite{Wyner_74, Zamir_02}. The fundamental idea is to assign each source vector to a bin from a set of $2^{H(X|Y)+\epsilon}$ known bins. The encoder, Alice, transmits the bin number to the decoder, thus encoding $X$ with rate $R=H(X|Y)+\epsilon$. The decoder looks for the source vector inside the described bin with help of the side information $Y$.

The efficiency of an information reconciliation protocol sending a sequence $C$ through the public channel to help Bob recover $X$ using side information $Y$ can be measured using a quality parameter $f$. If we allow $|\cdot|$ to stand for the length of a variable, $f$ is defined by:
%CAMBIO 2

\begin{equation}
f = \frac{|C|}{H(X|Y)} \geq 1
\label{eq:efficiency}
\end{equation}

According to this definition of efficiency, it takes its lowest value $f = 1$ in the optimal case, i.e. when the information published for reconciliation is the minimum possible information.

\subsection{Previous Work}
\label{sec:cascade}

Several protocols have been studied for information reconciliation. Many of them have been discussed in the context of Quantum Key Distribution (QKD) as it is one of the main scenarios of real secret key distillation. 

Brassard and Salvail proposed the Cascade protocol in \cite{Brassard_94} for binary variable reconciliation. Cascade despite being highly interactive remains the most widely used protocol. It offers to its advantage a simple description and a relatively low efficiency value. Other protocols include a protocol by Liu et al. \cite{Liu_03} that combines advantage distillation and information reconciliation, and Winnow \cite{Stucki_05}, a protocol in which Alice and Bob exchange the syndrome of a Hamming code for each block.

LDPC codes have been proposed for coding correlated sources in \cite{Muramatsu_05}, though no explicit codes were given. A rate adaptive contruction with non binary LDPC codes was proposed in \cite{Kasai_10}. On \cite{Elkouss_09} LDPC codes were optimised for the binary symmetric channel (BSC) and used to reconcile binary variables. The efficiency of the codes was close to $1$ for crossover probabilities near the codes' thresholds, however as only a discrete number of codes was available the efficiency exhibited a saw behaviour (see Fig.~\ref{fig:ex}). A rate adaptive protocol was proposed in \cite{Elkouss_10}, however the security of the protocol was not addressed and the impact of the excess of information on the public channel was not discussed.

\section{Rate Adaptive Information Reconciliation}
\label{sec:protocol}
\subsection{Formalism}
In this section we describe a protocol for the information reconciliation problem based on Wyner's coset scheme, briefly sketched above. Before describing the protocol we review some basic formalism. 

Let $\zeta(n, k)$ be a binary linear code of length $n$,  $k$ information symbols, and $R_0 = k/n$ its information rate. This code can be specified by a parity matrix $H$. Let $x$ be a $n$-length vector, such that $m(x) = H x^T$ stands for the syndrome of $x$. The code $\zeta(n, k)$ contains every $n$-length vector $v$ such that $m(v) = 0$. The best way to choose the bins for Wyner's schema, is to choose bins with a structure that allows differentiating between them. One natural way is to assign a bin to each coset of a linear code~\cite{Zamir_02}. Each bin can be seen as an affine code, characterised by syndrome $m_b$, that contains every $n$-length vector $v$ such that $m(v) = m_b$. There are $2^{n-k}$ different syndromes, thus allowing Alice to encode $x$ with rate $(n-k)/n$.

It was first shown in~\cite{Liveris_02} and generalised in~\cite{Muramatsu_05} that LDPC codes can be successfully used in order to address the problem of coding correlated sources with side information at the decoder. The message passing decoder must be modified to take into account the different syndromes and, channel coding techniques that lead to channel capacity approaching codes, lead also to codes approaching the Slepian-Wolf limit~\cite{Elkouss_09}. However, a linear code reveals a fixed amount of information independently of the channel characteristics which might not be appropriate in many situations. An scenario with changing statistics can arise in real settings due, for example, to the sensitivity of physical devices or to the presence of an active eavesdropper. To address the problem of secret key agreement when the statistics of the channel can vary from execution to execution, a suitable solution is provided by puncturing and shortening strategies (see Fig.~\ref{fig:ps}). 

A punctured code modifies an existing $\zeta(n, k)$ code by removal of a set of $p$ from the total $n$ symbols, thus becoming a code of length $n-p$ and dimension $k$, $\zeta'(n-p, k)$. In the same fashion, a shortened code is a modified code in which $s$ symbols from the code are known or fixed. A shortened code becomes a code of length $n-s$ and dimension $k-s$, $\zeta'(n-s, k-s)$. A code $\zeta(n, k)$ in which $p$ symbols are punctured and $s$ symbols are shortened becomes a code with rate:

\begin{figure*}[!t]
\centering
\subfloat[Puncturing]{\includegraphics[width=0.5\linewidth]{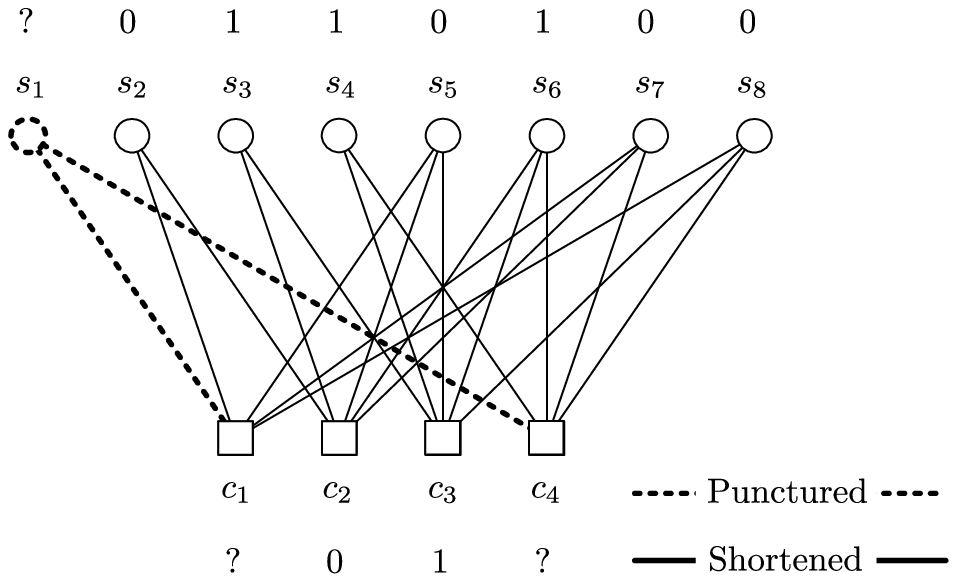}
\label{fig:puncturing}}
\hfil
\subfloat[Shortening]{\includegraphics[width=0.41\linewidth]{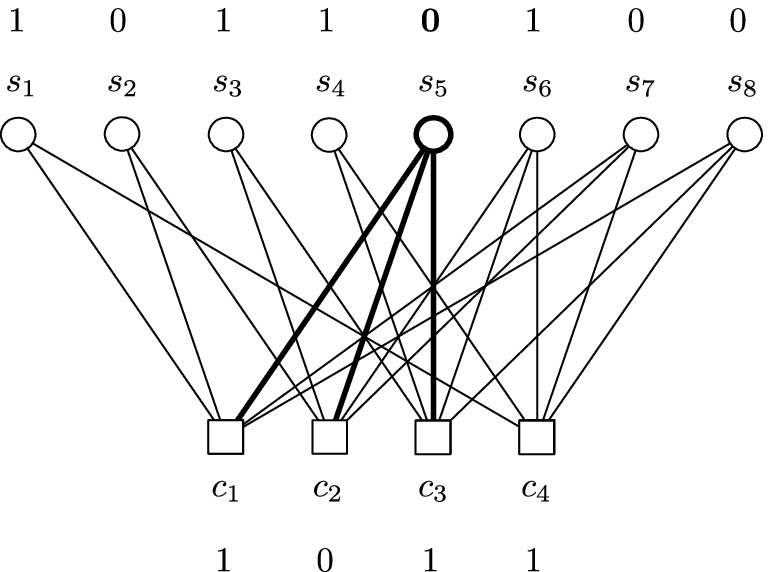}
\label{fig:shortening}}
\caption{Example of Tanner graph of an LDPC code with puncturing and shortening strategies applied on only one symbol.}
\label{fig:ps}
\end{figure*}

\begin{equation} 
R = \frac{k - s}{n - s - p} 
\label{eq:rate}
\end{equation}

This expression can also be written as a function of $R_0$, $\sigma = s/n$ and $\pi = p/n$: the original coding rate, the fraction of shortened symbols and the fraction of punctured symbols, respectively. Puncturing and shortening provide the means to adapt the rate of an existing code, however once chosen $p$ and $s$ the new rate is fixed. It should also be noted that there is a certain amount of efficiency loss as the percentage of punctured and shortened bits increases and even a limiting threshold of puncturing depending on the code \cite{Pishro-Nik_07}. 

\subsection{Rate Adaptive Protocol}
The following definition delineates a generic protocol able to adapt the information rate to varying channel parameters through puncturing and shortening strategies, $s+p$ random bits are added to the original strings. The protocol transmits $s+n-k$ bits through the public channel, which can stand for the code syndrome and $s$ shortened bits.  

\begin{definition}
Let $\zeta(n,k)$ be a linear code and $s,p \in \mathbb{N}$ two parameters such that $0 \leq s \leq k$, $0 \leq p \leq n$, $s+p\leq n$. An $sp$-protocol allows two parties holding $x$ and $y$ two $(n-p-s)$-length binary sequences to reconcile their strings. This protocol transmits $s+n-k$ bits through a public channel and extends both sequences $x$ and $y$ with $s+p$ random bits into $\hat{x}$ and $\hat{y}$ two $n$-length sequences.
\end{definition}

We now describe a practical $sp$-protocol which is a formal and simplified version of a protocol described in \cite{Elkouss_10} adapted for easier analysis. Let $R_0$ be the rate of $\zeta(n, k)$, in order to reconcile their string the two parties Alice and Bob perform the following steps:

\subsubsection*{Step 0}

Alice and Bob fix a parameter $\delta = \sigma + \pi$ standing for the number of symbols to either puncture or shorten, this allows them to reconcile the same amount of information on each protocol execution. They characterise as well $f(p_\mathrm{err})$, the efficiency function describing the behaviour of the code under shortening and puncturing, where $p_\mathrm{err}$ stands for the error probability.

Prior to the execution of the protocol Alice and Bob might have an estimate of the discrepancies between their strings or they might have published and subsequently discarded a subset of their original strings in order to infer the error probability $p_\mathrm{err}$. Once estimated $p_\mathrm{err}$ and having measured the quality of $\zeta$ under perforation and shortening, Alice and Bob choose $s$ and $p$ such that $R$ the rate of the equivalent code allows to reconcile both strings with high probability while minimising $s$.

\subsubsection*{Step 1}

\begin{figure}
\centering
\includegraphics[width=\linewidth]{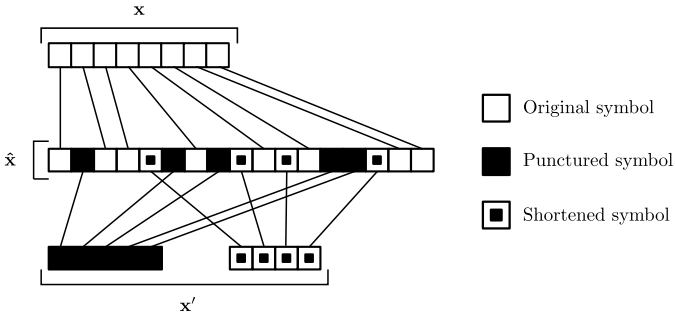}
\caption{Extended string construction. It is shown how the extended string $\hat{x}$ is constructed from a random permutation of two strings: the original string to be reconciled, $x$, and a string consisting of punctured and shortened symbols, $x'$.}
\label{fig:extended-string}
\end{figure}

Alice creates an extended string $\hat{x}$ (see Fig.~\ref{fig:extended-string}):

\begin{equation}
\hat{x} = g \left( x | r_{A}(p) | r_{A}(s) \right)
\end{equation}

\noindent where $g$ is a permutation of $x | r_A(p) | r_A(s)$, 
$r_A(p)$ is a random string of length $p$, and $r_A(s)$ is a random string of length $s$.

Alice transmits to Bob $m(\hat{x})$ and $r_A(s)$.

\subsubsection*{Step 2}

Bob receives Alice message and constructs an extended string $\hat{y}$:

\begin{equation}
\hat{y} = g \left( y | r_B(p) | r_A(s) \right)
\end{equation}

\noindent where $r_B(p)$ is a random string of length $p$ generated by Bob.

Bob recovers $\hat{x}$ with high probability using the modified belief propagation decoder described in \cite{Liveris_02}.

\begin{example}

Alice and Bob have a code $\zeta(2 \times 10^5, 10^5)$ with an empirical efficiency below $f(p_\mathrm{err}) \leq 1.09$ in the range $[0.065, 0.075]$ (see Fig.~\ref{fig:ex}) for $\delta = 0.05$. Alice transmits to Bob a string of length $1.9 \times 10^5$ over a BSC with known crossover probability $p_\mathrm{err} = 0.068$. In a BSC the conditional entropy can be expressed as $H(X|Y) = h(p_\mathrm{err})$, and thus the maximum coding rate $R = 1 - f(p_\mathrm{err}) h(p_\mathrm{err})$. Then, from Eq.~\ref{eq:rate}, they should puncture $p = 5,772$ bits and shorten $s=4,228$ bits to reconcile their extended strings with high probability and $f \leq 1.09$.

\end{example}

\vspace{2mm}

An important remark here is that Alice and Bob reconcile their \textit{extended} strings with efficiency $f$ close to $1$, while $f$, as defined on Eq.~\ref{eq:efficiency} for reconciling the \textit{original} strings, is higher. In the next section we show that the amount of distillable secret bits is not diminished by the higher $f$ value and, indeed, the relevant figure is the reconciliation efficiency of the extended strings.

\section{Security Analysis}
\label{sec:security}
The security of $sp$-protocols is addressed in this section. As a first step we review the privacy amplification results that allow to take into account the impact of reconciliation in the final key.

We introduce another entropy measure: min-entropy, as it is used in the following discussion. It is defined as:

\begin{equation}
H_\infty (X) = - \log \max_x P_X(x)
\end{equation}

Generally $H_\infty (X)\leq H(X)$, being equal only if $X$ outcomes are given by a uniform distribution. We further define the conditional min-entropy as:

\begin{equation}
H_\infty (X|Y) = \min_y H_\infty (X|Y=y)
\end{equation}

\begin{theorem}
\label{th:privacy-amplification}
Given three constants $\delta, \Delta_1, \Delta_2 \geq 0$, after $n$ uses of a binary symmetric channel ruled by $P_{Z'|X}$, if Eve's min-entropy on $X$ is known to be bounded as $H_\infty (X|Z'=z')\geq \delta n$, there exists (\cite{Maurer_00}) an extractor function $E:F_2^n\times F_2^u \rightarrow F_2^k$, with $u\leq \Delta_1 n$ and $k\geq (\delta -\Delta_2)n$, such that if Alice and Bob agree on secret key $K=E(X,U)$, where $U$ is a sequence of $u$ random uniform bits, the entropy of $K$ is given by:

\begin{equation}
\label{eq:hmin-redundancy}
H(K|U,Z'=z')\geq k - 2^{-n^{1/2-o(1)}}
\end{equation}
\end{theorem}

\noindent which wipes all the information from the eavesdropper provided that Alice and Bob can estimate $H_{\infty}(K|Z')$. 

The effects of the $|C|$ redundancy bits shared on the conditional min-entropy can also be bounded using a security parameter $t$ with probability $1-2^{-t}$ ~\cite{Maurer_00}:

\begin{equation}
\label{eq:hmin-z}
H_{\infty}(X|Z'=zc)\geq H_{\infty}(X|Z=z) - |C| - t
\end{equation}

\noindent measuring the interest of good information reconciliation, every redundancy bit used in this phase reduces the final secret key.

We proceed to demonstrate that the use of an $sp$-protocol does not impose any constraint on the achievable secret key rate. Moreover, from this demonstration it is possible to infer that the quality of the information reconciliation procedure depends only on the quality of the error correction code. We begin with the proof of the following lemma (Lemma~\ref{lem}) that allows to exploit the random construction of the punctured and shortened bits in the proposed protocol.

\begin{lemma}
\label{lem}

Let $X$, $Y$ and $Z$ be three random variables, if $Y$ is independent from variables $X$ and $Z$ the mutual min-entropy of $X$ and $Y$ conditioned to $Z$ can be expressed by:

\begin{equation}
H_\infty (XY|Z) = H_\infty (X|Z) + H_\infty (Y)
\end{equation}

\end{lemma}

\begin{proof}

\begin{equation}
H_\infty (XY|Z) = \min_{z} H_\infty (XY|Z=z)
\end{equation}

\begin{equation}
 = - \min_{z} \log \max_{xy} P(xy|z)
\end{equation}

\begin{equation}
\label{eq:pxpy}
 = - \min_{z} \log \max_{xy} P(x|z) P(y|z)
\end{equation}

\begin{equation}
= - \min_{z} \left[ \log \max_{x} P(x|z) + \log \max_{y} P(y|z) \right]
\end{equation}

\begin{equation}
\label{eq:endlemma}
 = H_\infty (X|Z) + H_\infty (Y)
\end{equation}

\noindent where Eq.~\ref{eq:pxpy} derives from the consideration that $X$ and $Y$ being independent variables, and Eq.~\ref{eq:endlemma} from $Y$ and $Z$ being independent variables.

\end{proof}

\begin{theorem}
\label{th:minEnt}

Given a code $\zeta (n,k)$, a security constant $t$, the public communication $C$, and $Z$ the eavesdropper information, then the min-entropy of the variable $\hat{X}$ constructed by the $sp$-protocol, is with probability $1-2^{-t}$ greater or equal than that of using an adapted error correcting code of rate $R$ to reconcile $X$ and $Y$ minus the security constant:

\begin{equation}
H_\infty (\hat{X}|ZC) \geq H_\infty (X|Z) - |X|(1-R) - t
\end{equation}

\end{theorem}

\begin{proof}

Directly given by Eq.~\ref{eq:hmin-z}:

\begin{equation}
H_\infty (\hat{X}|ZC) \geq H_\infty (\hat{X}|Z) - |C| - t  
\end{equation}

Distinguishing in $\hat{X}$ part of the variable that corresponds to the sequence to be reconciled, $X$, and the additional variable used to extend the original sequence, $X'$ (see its correspondence with strings in Fig.~\ref{fig:extended-string}):

\begin{equation}
= H_\infty (XX'|Z) - |C| - t
\end{equation}

Since $X'$ is independent of $Z$ and $X$ by construction, Lemma~\ref{lem} can be applied:

\begin{equation}
= H_\infty (X|Z) + H_\infty (X') - |C| - t
\end{equation}

The entropy of $H_\infty (X')$ takes the value of the number of random $p + s$ bits:

\begin{equation}
= H_\infty (X|Z) + |X| \frac{\pi + \sigma}{1 - \pi - \sigma} - |C| - t
\end{equation}

The length of the conversation $|C|$ is $s + n - k$, which in the proposed protocol stand for the $s$ shortened bits and the syndrome of $X'$. It can be written as a function of the size of $X$, $\pi$ and $\sigma$:

\begin{equation}
= H_\infty (X|Z) + |X| \frac{\pi + \sigma}{1 - \pi - \sigma} - |X|  \frac{(1 - R_0) + \sigma}{1 - \pi - \sigma} - t
\end{equation}

and thus

\begin{equation}
= H_\infty (X|Z) - |X|  (1 - R) - t
\end{equation}

\end{proof}

\section{Numerical Results}
\label{sec:num}
\begin{figure}
\centering
\includegraphics[width=\linewidth]{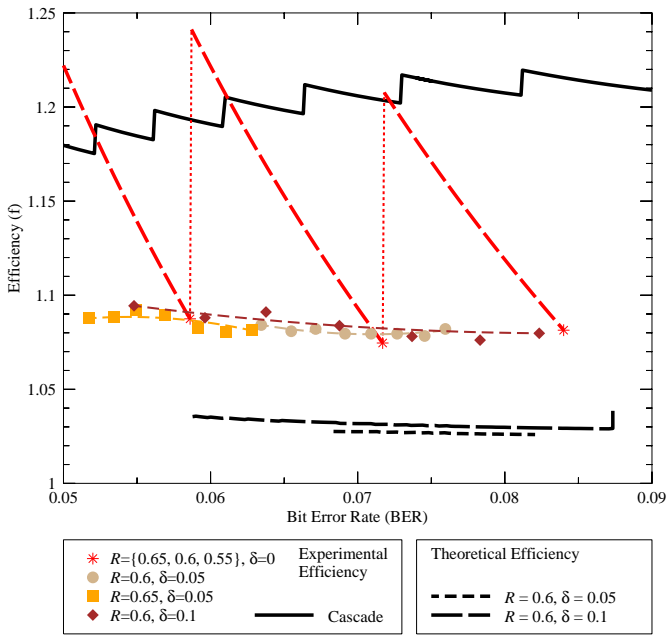}
\caption{Reconciliation efficiency of Cascade~\cite{Brassard_94}, LDPC codes without puncturing and shortening strategies~\cite{Elkouss_09}, and the $sp$-protocol in a practical setting as defined in Eq.~\ref{eq:efficiency}. Two LDPC codes have been chosen to cover the crossover range $p_\mathrm{err} \in [0.055, 0.08]$ using the proposed $sp$-protocol. Both codes, $\zeta_1 (2 \times 10^5, 1.2 \times 10^5)$ with coding rate $R = 0.6$ and $\zeta_2 (2 \times 10^5, 1.3 \times 10^5)$ with coding rate $R = 0.65$, allow to cover the range with $\delta = 0.05$, while $\zeta_2$ with $\delta = 0.1$ also covers the range of interest. A third code with rate $R = 0.55$ has been used in order to compare the efficiency of the studied crossover range with a direct strategy, i.e. without using puncturing or shortening, as proposed in~\cite{Elkouss_10}.}
\label{fig:ex}
\end{figure}

We discuss the efficiency of several protocols in this section. In order to illustrate the performance of the $sp$-protocol in Fig.~\ref{fig:ex} we compare the results of adapted LDPC codes to regular LDPC codes without adaptation and to Cascade. We show as well the theoretical efficiency in case of infinite length~\cite{Elkouss_10}, this curve indicates the expected asymptotic behaviour of the protocol. 

Following Theorem~\ref{th:minEnt} two strings can be reconciled with the efficiency of a rate adapted code. In the figure, the efficiency of the punctured and shortened codes is below $1.1$ in the whole range of $p_\mathrm{err}$, close to the theoretical limit. In comparison the codes without adaptation offer a better result close to their threshold but the efficiency quickly drops as the working point moves away from the threshold. On the other hand Cascade exhibits a poorer efficiency on the $p_\mathrm{err}$ range considered.

\section{Conclusion}

On this paper it has been discussed the problem of information reconciliation in the context of secret key agreement. The $sp$-protocol, a simple protocol based on puncturing and shortening LDPC codes has been proposed. This protocol allows the eavesdropper to gather the same amount of information than an adapted code would reveal; even if it is exchanged more data on the public channel. 

It had been argued that information reconciliation based on error correction codes was not optimal for channels with changing characteristics~\cite{Elkouss_09}, having Alice and Bob access to a discrete set of codes the efficiency of the reconciliation exhibits a saw behavior.  The $sp$-protocol allow Alice and Bob to reconcile their chains with a continuous efficiency curve, and as the efficiency of LDPC codes under puncturing and shortening can be analytically described and optimised, the results proved in this paper allow to address the information reconciliation problem as a code design problem. The numerical data on section \ref{sec:num} indicate that efficiency values close to the theoretical limits can be obtained. 

\section*{Acknowledgment}

This work has been partially supported by the project Quantum Information Technologies in Madrid\footnote{http://www.quitemad.org} (QUITEMAD), Project P2009/ESP-1594, \textit{Comunidad Aut\'onoma de Madrid}.

The authors would like to thank the assistance and computation resources provided by \textit{Centro de Supercomputaci\'on y Visualizaci\'on de Madrid}\footnote{http://www.cesvima.upm.es} (CeSViMa).

\bibliographystyle{IEEEtran}
\bibliography{isita2010}

\vfill

\end{document}